\documentclass[prl,twocolumn,groupedaddress,showpacs]{revtex4b4}
\usepackage{graphicx}
\begin{document}

\title{Polytetrahedral Clusters}

\author{Jonathan P. K. Doye}
\email{jon@clust.ch.cam.ac.uk}
\affiliation{University Chemical Laboratory, Lensfield Road, Cambridge CB2 1EW, United Kingdom}
\author{David J. Wales}
\affiliation{University Chemical Laboratory, Lensfield Road, Cambridge CB2 1EW, United Kingdom}
\date{\today}
\begin{abstract}
By studying the structures of clusters bound by a model potential that favours polytetrahedral 
order, we find a previously unknown series of `magic numbers' (i.e.\ sizes of special
stability) whose polytetrahedral structures
are characterized by disclination networks that are analogous to hydrocarbons.
\end{abstract}

\maketitle

Polytetrahedral order \cite{NelsonS,Sadoc99} has become an increasingly important concept in condensed-matter physics.
Such polytetrahedral structures, for which the whole of space can be naturally divided up into
tetrahedra with atoms at their vertices, are the basis of crystalline Frank-Kasper 
phases \cite{FrankK58,Shoemaker},
and have been invoked in order to understand the structure of quasicrystals \cite{Elser85,Sachdev85b,Audier86} 
and atomic liquids and glasses \cite{NelsonS,Nelson83a,Doye96a}.
However, little is known about the consequences of polytetrahedral order for 
the structure of clusters and nanoparticles. 
This situation contrasts with close-packing schemes, which give rise to 
fascinating cluster structures, such as
Mackay icosahedra \cite{Mackay}, Marks decahedra \cite{Marks84} and Leary tetrahedra \cite{Leary99}.

However, recent experiments indicate that small cobalt clusters can have polytetrahedral order \cite{Dassenoy00}, 
although only limited information about their detailed structure could be obtained.
Furthermore, there is an increasing interest in mixed metal 
clusters \cite{Cozzini96,Hodak00}, 
and polytetrahedral structures would be expected for those alloys that exhibit 
Frank-Kasper or quasicrystalline phases in bulk.

The distinctive features of polytetrahedral packings stem from the inability of regular tetrahedra 
to fill all space. When five regular tetrahedra are packed around a common edge there is 
a small angular deficit of $7.4^\circ$, whose closure requires a small distortion of the tetrahedra.
If this method of packing is extended to larger collections of tetrahedra, local icosahedral coordination
results, but the strain that needs to be introduced to close all the 
gaps grows very rapidly. Therefore, in order to form an extended 
polytetrahedral structure, sites where six tetrahedra share a common edge need to be introduced---a 
negative disclination line is said to run along this common edge. 
Even though the local distortion required to remove the overlap that occurs when packing
six regualar tetrahedra is larger, the overall strain is reduced.  

Polytetrahedral packings can therefore be described by a network of disclination 
lines threading an icosahedrally-coordinated medium.
In Frank-Kasper crystals this disclination network is ordered and periodic, whereas it
has been suggested that atomic liquids and glasses are characterized by disordered 
entangled disclination networks \cite{Nelson83a,NelsonS}.

\begin{figure}
\includegraphics[width=8.2cm]{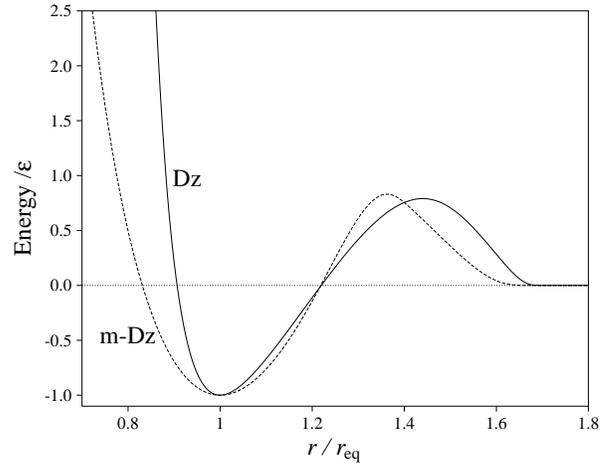}
\caption{Comparison of the Dzugutov potential in its original (Dz) and modified (m-Dz) form. 
}
\label{fig:potentials}
\end{figure}

Many very small clusters naturally form polytetrahedral clusters. 
The 13-atom icosahedron (20 slightly distorted tetrahedra sharing a vertex)
is an extremely common structure for rare gases \cite{Echt81}, metal \cite{Martin96}
and molecular \cite{Martin93} clusters and is generally favoured over a close-packed structure 
because of its lower surface energy. However, structures that continue this polytetrahedral
packing soon become disfavoured for most materials because the associated strain cannot
be accommodated. The largest polytetrahedral clusters have been obtained for a model system
where the width of the potential allows the system to tolerate ordered polytetrahedral 
structures up to $N$$\approx$70 \cite{Doye97d}.

\begin{figure*}
\includegraphics[width=18cm]{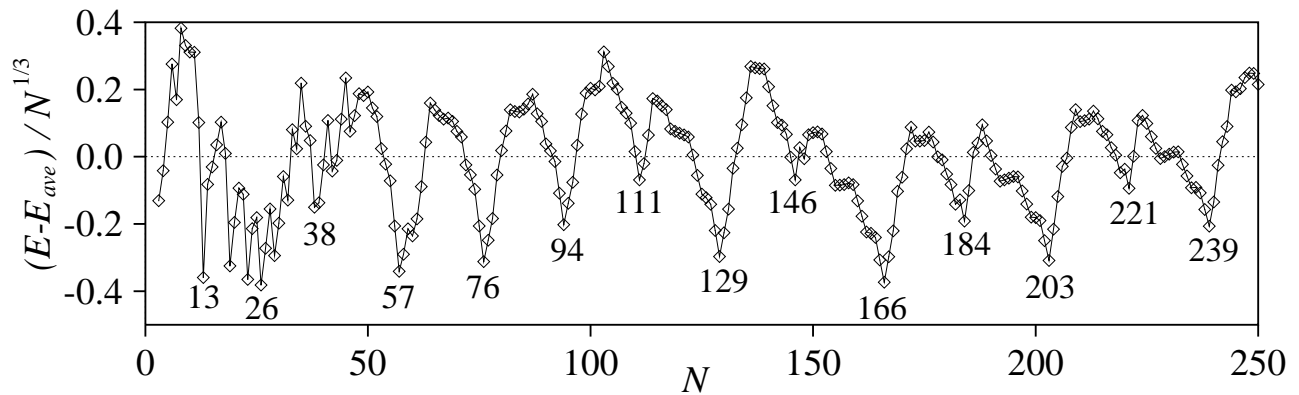}
\caption{Energies of the lowest-energy minima as a function of size relative to $E_{\rm ave}$,
where $E_{\rm ave}$ is a four parameter fit to these energies. 
$E_{\rm ave}=-3.040N+2.023N^{2/3}+1.949N^{1/3}+0.545$. }
\label{fig:EvN}
\end{figure*}

A model system that exhibits polytetrahedral clusters and enables us to make structural predictions 
about such clusters would therefore be of great interest.
Here we seek to address this issue
by studying clusters interacting with a potential of the Dzugutov form \cite{Dzugutov91}: 
\begin{eqnarray}
V(r)&=&A(r^{-m}-B) \exp\left( {c\over r-a}\right) \Theta(a-r) + \nonumber \\
    & &B \exp\left({d\over r-b}\right) \Theta(b-r), 
\end{eqnarray} 
where $\Theta(x)$, the Heaviside step function, is 0 for $x<0$ and 1 otherwise. 
The total potential energy of a cluster is then $E=\sum_{i<j} 
V(r_{ij})$, where $r_{ij}$ is the distance between atoms $i$ and $j$.
This potential was designed to encourage polytetrahedral and local icosahedral order 
in supercooled liquids through the introduction of a local maximum in the potential near
to $\sqrt 2$ times the equilibrium pair distance
(Fig.\ \ref{fig:potentials}) that disfavours close-packed structures \cite{Dzugutov93b}.
This maximum also somewhat resembles the first of the Friedel oscillations
that can occur for metal potentials.
Interestingly, under certain conditions this potential is able to form a dodecagonal 
quasicrystal \cite{Dzugutov93}.

For the original parameterization of the Dzugutov potential, clusters form non-compact polytetrahedral
structures composed of needles, rings and three-dimensional networks of interpenetrating
and face-sharing icosahedra \cite{Doye01a}. Thus the original potential cannot 
provide a realistic model of the compact polytetrahedral clusters formed for cobalt
or that might occur for metallic alloys. 
Non-compact structures occur because the relatively narrow potential well (Fig.\ \ref{fig:potentials})
does not allow the system to accommodate the strain associated with compact polytetrahedral clusters.
Therefore, we chose a new parameterization of the Dzugutov potential that both increases the width of 
the minimum and reduces the width of the potential maximum.
As envisaged, this new potential gives rise to compact polytetrahedral clusters.
The parameters in our modified potential have the values
\begin{equation}
\begin{array}{*{3}{c@{\quad}}c}
A=3.00  & B=2.109 & a=1.65 & b=1.94 \\
c=0.52  & d=0.55 & m=4.
\end{array}
\end{equation} 
The pair potential has 
a maximum at $r_{\rm max}$=
1.36$\,r_{\rm eq}$ of height 
$0.83\epsilon$, 
where $r_{\rm eq}$ and $\epsilon$ are the equilibrium pair separation and well depth, respectively.

To find the global minima of clusters interacting with this potential we used
the basin-hopping algorithm \cite{Li87a,WalesD97}. For each size up to $N$=100 we
performed five runs of $100\,000$ steps starting from a random configuration. 
We also performed short runs starting from configurations generated 
by adding or removing the appropriate number of atoms from some of the lowest-energy
minima for sizes one, two or three atoms above or below the current size.
These latter seeded runs were repeated until no new putative global minima were found
and were particularly important because the roughness of the energy landscape \cite{Doye01a} 
makes optimization from a random starting point particularly difficult.
Indeed above $N$=100 no runs from a random starting point found the global minima, so 
in this size range we used a different approach. Using the structural principles
obtained from the particularly stable structures for clusters with less than 100 atoms,
we were able to construct a series of candidate geometries for particularly
stable sizes in the range $N$=100--250. 
These structures then served as the initial seed configurations for a series 
of short basin-hopping runs for the intervening sizes.
Again seeded runs using the lowest-energy minima of nearby sizes were repeated until no 
further improvements were obtained for any size.

The energies of the resulting putative global minima are depicted in 
Fig.\ \ref{fig:EvN} in a manner that emphasises particularly stable minima 
or `magic numbers'. At small sizes the magic numbers ($N$=$13, 19, 23, 26, 29$) are those 
expected for polytetrahedral growth upon the 13-atom icosahedron and are similar to
those seen for Lennard-Jones clusters \cite{Northby87} and their experimental analogue, 
argon clusters \cite{Harris84}. The next magic number corresponds to the disc-like 38-atom cluster, which was 
previously found for clusters interacting with the original Dzugutov potential and which has 
a single disclination line running along its axis \cite{Doye01a}.
Then, for the rest of the size range we consider, there are series of roughly equally spaced 
minima in Fig.\ \ref{fig:EvN} that correspond to a new sequence of magic numbers.

\begin{figure}
\includegraphics[width=8.2cm]{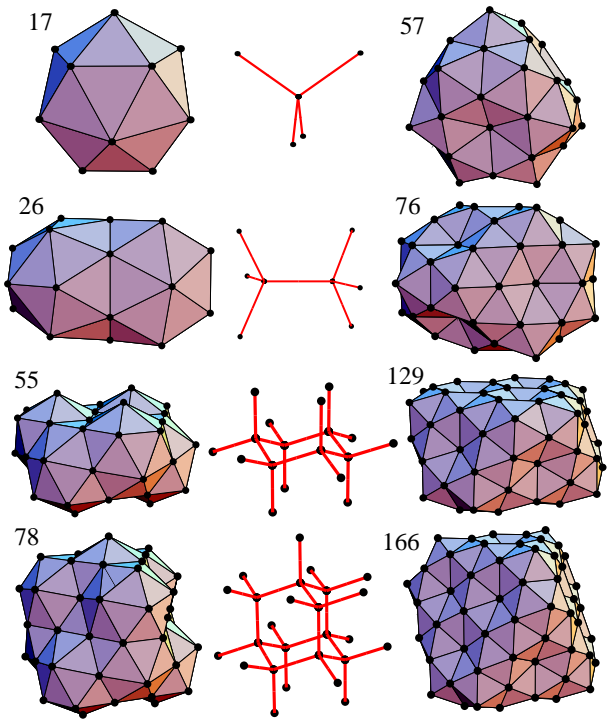}
\caption{Structures of some of the magic number clusters. 
On the right is the complete structure, in the middle the corresponding disclination
network and on the left the structure that is at the centre of the cluster.  
All three have the same orientation.}
\label{fig:aMdis}
\end{figure}

Some of the structures of these new clusters are depicted in the right-hand column 
of Fig.\ \ref{fig:aMdis}. Each cluster consists of a disclinated central structure that 
is surrounded by an overlayer in which an atom is added to each face and 
above each vertex that is not at the end of a disclination line.
This is the same overlayer as for the initial polytetrahedral growth on the 13-atom 
icosahedron, and in that context has been called the anti-Mackay \cite{Doye97d} 
or face-capping \cite{Northby87} overlayer.  
This overlayer does not extend the disclinations of the central cluster, thus giving rise to
characteristic six-fold pits where the disclinations exit onto the surface of 
the resultant cluster (Fig.\ \ref{fig:aMdis}).

\begin{figure}
\includegraphics[width=8.2cm]{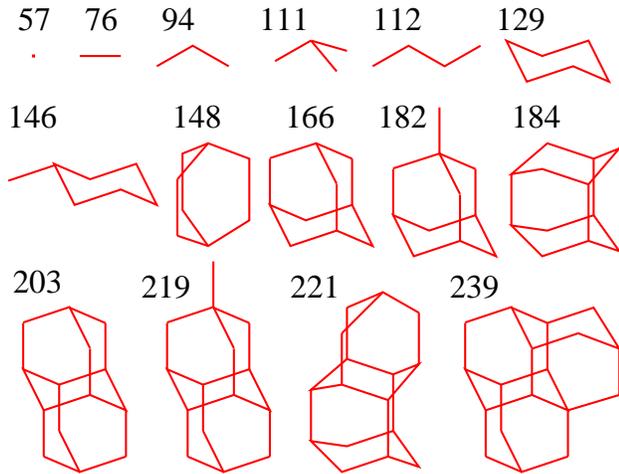}
\caption{Schematic representations of the disclination networks for the particularly stable clusters. 
The terminal disclination lines have been omitted for clarity.
The networks, thus drawn, are analogous to the carbon backbones of a series of hydrocarbons.}
\label{fig:alkanes}
\end{figure}

At the centre of the 57-atom cluster is a 17-atom structure in which the central atom is the node
for four disclinations in a tetrahedral arrangement. 
This coordination shell is commonly found in the Frank-Kasper phases \cite{Shoemaker}.
Furthermore, the atoms of all the central clusters either have this or 
an icosahedral coordination shell, thus giving rise to a tetrahedrally
coordinated disclination network (Fig.\ \ref{fig:alkanes}). 
There are two bulk Frank-Kasper phases that involve such networks,
the C14 and C15 phases. In the C15 phase the disclination network
has the structure of the diamond lattice and in the C14 phase the
wurtzite structure. All the cluster centres can be considered to be
fragments of these two Frank-Kasper phases, except the centre of the 
221-atom cluster which involves a mixture of the two phases.

The easiest way to understand the progression of structures is to note the
correspondence between the disclination networks and hydrocarbon structures (Fig.\ \ref{fig:alkanes}). 
The disclination network of the 57-atom cluster is analogous to methane.
Next comes a series corresponding to the linear alkanes, ethane (76), propane (94) and n-butane (112),
and the branched alkane, isobutane (111). At this point it becomes favourable
to form more compact structures analogous to cycloalkanes, e.g.\ the chair form of cyclohexane (129) 
and methyl-cyclohexane (146). 
Above this size structures analogous to cage hydrocarbons \cite{Olah} are favourable, 
such as bicyclo[2.2.2]octane (148), adamantane (166), diamantane (203) and 
triamantane (239). 
It is noticeable that the most stable of these latter magic numbers correspond to
the polymantanes (or diamondoids), where the central structures are fragments of the C15 phase.
C14 and mixed disclination networks are only competitive in between these sizes 
when they are competing with structures analogous to methyl-polymantanes.

The analogy to the hydrocarbons also allows ready prediction of the structure of larger clusters. 
For example it is well-known that the next polymantane, tetramantane has three isomers \cite{Olah}.
However, as for the original parameterization of the Dzugutov potential \cite{Roth00}, 
the body-centred-cubic (bcc) lattice is lowest in energy for bulk. 
A comparison of $E_{\rm ave}$ to a similar function fitted to a series of bcc rhombic dodecahedra
indicates that bcc clusters are lowest in energy beyond $N$$\approx$1400. 

Experiments can usually only probe cluster structure indirectly and so comparisons with 
calculations from candidate geometries are required for structural identification. 
Therefore, to enable the potential identification of the structures described here,
all are accessible from the Cambridge Cluster Database \cite{Web}.
Furthermore, as the series of magic numbers does not coincide with any previously known \cite{Martin96},
there is also the potential for identification through mass spectral abundances.

Calculations for an example cluster, the 166-atom adamantane analogue,
are consistent with the experimental results for cobalt clusters \cite{Dassenoy00}. 
The calculated scattering function reproduces many of the features observed in the x-ray data, 
and the structure has perpendicular planes of atoms that can account for the 
square arrays of lattice fringes seen in the electron microscopy images.
However, this agreement may result more from the polytetrahedral order than 
the actual detailed structure of our clusters \cite{Dassenoy00}.

It is remarkable that many of the unusual cluster structures that have been 
predicted by theory have been subsequently observed experimentally.
For example, the Mackay icosahedron \cite{Mackay}, first suggested in 1962, 
has since been seen for a wide variety of systems over a large size range \cite{Martin96,Miehle89}.
Furthermore, the small truncated octahedron at $N$=38 and Marks decahedron at $N$=75, 
whose stability was first identified in calculations on model clusters \cite{Doye95c}, 
have since been identified for Ni$_{38}$ \cite{Parks97}, Au$_{38}$ \cite{Alvarez97}, 
and Au$_{75}$ \cite{Cleveland97b}.
Most recently the Leary tetrahedron, a surprise global minimum for a 98-atom 
Lennard-Jones cluster \cite{Leary99}, has since been found for (C$_{60}$)$_{98}$ \cite{Branz00}.
Therefore, it would be no surprise if the polytetrahedral structures that we have 
described here were likewise to be positively identified experimentally.

JPKD is grateful to Emmanuel College, Cambridge for financial support.

\end{document}